\PassOptionsToPackage{table}{xcolor}
\documentclass{svproc}
\usepackage{graphicx}%
\usepackage{multirow}%
\usepackage{booktabs}%
\usepackage{mathrsfs}%
\usepackage[title]{appendix}%
\usepackage{amsmath,amssymb,amsfonts}
\usepackage{xcolor}%
\usepackage{textcomp}%
\usepackage{manyfoot}%
\usepackage{booktabs}%
\usepackage{algorithm}%
\usepackage{algorithmicx}%
\usepackage{algpseudocode}%
\usepackage{listings}%
\usepackage[compatibility=false]{caption}
\usepackage{subcaption}
\usepackage{bm}
\usepackage{tikz}
\usetikzlibrary {arrows.meta,bending,positioning}
\usetikzlibrary{patterns}
\usepackage{matlab-prettifier}
\usetikzlibrary{shapes.multipart}
\usetikzlibrary{tikzmark}
\usepackage[utf8]{inputenc}
\usepackage{flushend}
\usepackage{pgfplots}
\usepackage{blkarray}
\usetikzlibrary{calc}
\usetikzlibrary{decorations.pathreplacing,angles,quotes}
\usetikzlibrary {decorations.fractals,spy}
\usepackage{tikz-dimline}
\usepackage{makecell}
\usepackage{mwe}
\newcolumntype{C}[1]{>{\centering\arraybackslash}m{#1}}
\usepackage[table]{xcolor}
\definecolor{white}{rgb}{1,1,1}
\definecolor{NavyBlue}{rgb}{0, 0.2, 0.4}
\definecolor{Teal}{rgb}{0, 0.5, 0.5}
\definecolor{DarkGreen}{rgb}{0, 0.3, 0}
\definecolor{ForestGreen}{rgb}{0, 0.4, 0.1}
\definecolor{LimeGreen}{rgb}{0.5, 1.0, 0.5}
\definecolor{DarkOrange}{rgb}{1.0, 0.55, 0}
\definecolor{Peach}{rgb}{1.0, 0.8, 0.6}
\definecolor{Maroon}{rgb}{0.5, 0, 0}
\definecolor{Burgundy}{rgb}{0.4, 0, 0.1}
\definecolor{HotPink}{rgb}{1.0, 0.4, 0.7}
\definecolor{DeepPink}{rgb}{0.8, 0.1, 0.4}
\definecolor{Indigo}{rgb}{0.29, 0, 0.51}
\definecolor{SlateGray}{rgb}{0.44, 0.5, 0.56}
\definecolor{LightGray}{rgb}{0.8, 0.8, 0.8}
\definecolor{SteelBlue}{rgb}{0.27, 0.51, 0.71}
\definecolor{Turquoise}{rgb}{0.25, 0.88, 0.82}
\definecolor{olivegreen}{rgb}{0.75, 0.75, 0}
\definecolor{greenyellow}{rgb}{0.75, 1, 0}
\definecolor{violet}{rgb}{0.4940, 0.1840, 0.5560}
\definecolor{yellowochre}{rgb}{0.9290, 0.6940, 0.1250}
\definecolor{redorange}{rgb}{0.8500, 0.3250, 0.0980}
\definecolor{brownred}{rgb}{0.6350, 0.0780, 0.1840}
\definecolor{skyblue}{rgb}{0.3010, 0.7450, 0.9330}
\definecolor{shafgreen}{rgb}{0.4660, 0.6740, 0.1880}
\definecolor{red}{rgb}{1,0,0}
\definecolor{blue}{rgb}{0,0,1}
\definecolor{green}{rgb}{0,1,0}
\definecolor{yellow}{rgb}{1,1,0}
\definecolor{magenta}{rgb}{1,0,1}
\definecolor{cyan}{rgb}{0,1,1}
\definecolor{black}{rgb}{0,0,0}
\definecolor{gray}{rgb}{0.6,0.6,0.6}

\pgfplotsset{compat=1.14}

\usepackage{url}

\begin{document}
	\mainmatter 
	
	\title{Topology optimization of multimaterial aircraft pylons using generalized shape function approach}
	
	\titlerunning{MMTO of Aircraft Pylon} 
	
	\author{Swagatam Islam Sarkar \and Prabhat Kumar\inst{*}}
	
	\authorrunning{S.I. Sarkar and P. Kumar}
	
	\tocauthor{Swagatam Islam Sarkar and Prabhat Kumar}
	
	\institute{Indian Institute of Technology Hyderabad, Kandi, Telangana 502284, India \\ \inst{*}\url{pkumar@mae.iith.ac.in}}
	
	\maketitle

\begin{abstract}
As the primary structural component connecting the engine to the wing or fuselage, an aircraft pylon requires optimized structural efficiency; this paper provides topology optimization of multimaterial pylons using the  generalized shape function (gSF) approach. The gSF method uses $n$ natural-coordinate design variables per element to provide  optimized designs up to $2^n$ distinct material phases while promoting close to discrete material layouts in conjunction with the density and formulated Heaviside projection filters. Pylon structural compliance is minimized subject to volume constraints. Exploiting the geometric features of a typical pylon structure, multimaterial evolution is performed on a corresponding 2D design domain representing the midplane, with up to 14 candidate materials. The optimized two-dimensional layout is then extruded to achieve the corresponding three-dimensional optimized pylon structure. The Method of Moving Asymptotes is employed to achieve the final design variables. The resulting convergence histories exhibit smooth and stable objective minimization. The results highlight the capability of the multimaterial topology optimization framework to effectively optimized aircraft pylons with multiple candidate materials, without requiring a considerable expansion of the design variable set.

\end{abstract}
\section{Introduction}\label{sec1}
An aircraft pylon is an essential structural component that supports the engine and connects it to the aircraft wing. It transfers the loads produced by the engine and aerodynamic forces to the main airframe structure~\cite{stefanovic2021structural,kim2022structural}. During operation, it experiences combined effects from engine weight, thrust, aerodynamic loading, and dynamic flight conditions~\cite{kim2022structural,zettel2025jet}. The pylon must therefore achieve sufficient stiffness and strength while meeting given mass/resource requirements. This makes pylon design an attractive candidate for the systematic design optimization technique, structural/topology optimization (TO).  As an aerospace load-bearing substructure, the engine pylon presents a challenging TO problem because concentrated thrust, inertial, and gust loads must be transferred to the wing box while satisfying tight stiffness, fatigue, and mass constraints.

TO determines the material layout within a prescribed design domain by optimizing a specified objective while satisfying loading, boundary, and geometric constraints. Design domains are discretized using finite elements (FEs), including standard or advanced formulations~\cite{kumar2023honeytop90}. In single-material topology optimization, one design variable is assigned to each finite element. Its foundations were established by Bends{\o}e and Kikuchi through a homogenization-based formulation~\cite{bendsoe1988generating}, followed by the SIMP method~\cite{bendsoe1989optimal}. Since then, TO has become a powerful design tool for lightweight, high-performance structures across mechanical, civil, and aerospace engineering, owing to its ability to explore a design space far broader than conventional shape and size optimization~\cite{singh2025normalized}. Within aerospace applications, pylon-specific TO has an established track record. Remouchamps et al.~\cite{remouchamps2011application} applied a bi-level TO scheme to two Airbus pylons, showing that TO-informed layouts diverge from traditional designs, while Coniglio et al.~\cite{coniglio2019engine} extended pylon TO to jointly consider mass, stress, and engine thrust-specific fuel consumption, addressing non-conforming mesh interfaces. However, most pylon-focused TO studies assume a single, homogeneous material, whereas modern pylons increasingly combine materials of differing stiffness, density, and fatigue behavior to achieve further mass savings~\cite{khalid2021substituting}.

The multimaterial topology optimization (MMTO) simultaneously determines material layout with different phases~\cite{kumar2022topologyMM,sarkar2026generalized}. However, increasing the number of candidate materials, conventional approaches, including extended SIMP~\cite{gao2011mass}, Discrete Material Optimization (DMO)~\cite{stegmann2005discrete}, phase-field~\cite{tavakoli2014alternating}, and neural-network-based methods~\cite{chandrasekhar2021multi}, to name a few, often require increasing design variables and careful parameter tuning to suppress intermediate density interface regions~\cite{sarkar2026generalized}. Although strategies such as the peak-function-based formulation~\cite{yin2001topology}, ordered SIMP~\cite{zuo2017multi}, and sequential two-material decomposition~\cite{yang2018discrete} reduce the number of design variables, they require substantial parameter tuning as the number of candidate materials increases.

The generalized shape function (gSF) method extends one-dimensional, two-dimensional, and three-dimensional shape functions to a generalized $n$-linear form, mapping a compact set of natural coordinates $(\boldsymbol{\zeta} \in [-1,\,1]^n)$ onto the multimaterial simplex as design variables. This construction has the property of barycentricity, ensuring physically valid material fractions and yielding $2^n$ material phases with only $n$ design variables. In conjunction with classical filtering and the formulated projection scheme, the gSF formulation ensures near-discrete, 0-1-optimized solutions. We employ this method here to optimize multimaterial aircraft pylons using TO. 

The remainder of the paper is structured as follows. Section~\ref{sec2} provides the gSF formulation in brief. Section~\ref{sec3} presents the optimization formulation. The Pylon domain, loads, boundary conditions, and the multimaterial-optimized results for the pylon are described in Section~\ref{sec4}. Section~\ref{sec5} provides the concluding remarks.

\section{Multimaterial model with the gSF approach}\label{sec2}
This section summarizes the gSF approach~\cite{sarkar2026generalized} for multimaterial modeling, which is adopted in the paper. In the gSF framework with $n$D material element the density of any material $m$ of an FE can be calculated as~\cite{sarkar2026generalized}:
\begin{equation}\label{Density}
\begin{split}
    \rho_m =\prod_{i=1}^{n}\frac{(1+x_{im}\zeta_i)}{2}.
\end{split}
\end{equation}
Here, $n$ denotes the number of design variables, and $x_{im}$ is the coordinates of the $m$-th vertex of an $n$D hypercube of two unit side length, where the centroid is at the origin. It is well known that a density filter utterly controls the minimum feature size and prevents common numerical instabilities in optimized designs in a TO setting. And as demonstrated in~\cite{sarkar2026generalized}, applying classical density filtering on the design variable is equivalent to directly applying the same on the material density~\cite{sarkar2026generalized}. The filtered density corresponding to material $m$ of an element is evaluated as~\cite{sarkar2026generalized}: 
\begin{equation}\label{Filtered_density1}
    \begin{split}
        \tilde{\rho}_m &=\prod_{i=1}^{n}\frac{(1+x_{im}\tilde{\zeta}_i)}{2}.
    \end{split}
\end{equation}
The $i$-{th} filtered design variable of element $j$ is determined as~\cite{sarkar2026generalized}:
\begin{equation} \label{Filtering}
	\tilde{\zeta}_{ij} = \frac{\sum_{k} \hat{W}_{jk} \zeta_{ik}}{\sum_{k} \hat{W}_{jk}},\qquad (k \in N_f).
\end{equation}
Here, $\zeta_{ik}$ denotes the $i$-th design variable associated with the element $k$, while $N_f$ represents the set of neighboring elements within $r_\mathrm{min}$ (filter radius). $\hat{W}_{jk}$, a weighting factor, is evaluated by $\hat{W}_{jk} = \max({0,r_\mathrm{min}-d(j,k)})$. $d(j,k)$ denotes the centre-to-centre distance of the element $j$ and the element $k$. Filtering material densities yields solutions that deviate from a binary distribution of optimized densities. Consequently, a modified projection-based filter is adopted~\cite{sarkar2026generalized} and implemented to achieve crisp, discrete, optimized solutions. The projected density of material $m$ of an element is determined as:
\begin{equation} \label{Projected_filtered_density}
\bar{\rho}_{m}=\prod_{i=1}^{n}\frac{(1+x_{im}\bar{\zeta}_{i})}{2},
\end{equation}
where $\bar{\rho}_{m}$ is the projected density, i.e., the physical density of the $m$-{th} material within an element. $\bar{\zeta}_{i}$ represents the $i$-{th} projected design variable of an element and is evaluated as follows~\cite{sarkar2026generalized}:
\begin{equation} \label{Projection}
    \begin{split}
	\bar{\zeta}_{i}=2\left(\frac{\tanh \beta\gamma+\tanh \beta((\tilde{\zeta}_{i}+1)/2-\gamma)}{\tanh \beta\gamma+\tanh \beta(1-\gamma)}\right)-1
    \end{split}
\end{equation}
where $\beta$ and $\gamma$ are projection parameters that control the discreteness and smoothness of the design, respectively. $\tilde{\zeta}_{i}$ represents the $i$-{th} filtered design variable of an element. Readers can refer to Ref.~\cite{sarkar2026generalized} for a complete description of the gSF approach. Lastly, to quantify the near-binary solution of the optimized designs, the non-discreteness ($M_{\mathrm{nd}}$) in the gSF framework is determined as described in~\cite {sarkar2026generalized}. A value of $M_{\mathrm{nd}}$ approaching zero indicates that the material distribution is nearly discrete, or crisp, throughout the domain.

\section{Optimization Formulation}\label{sec3}
The modified SIMP method is reconstructed to incorporate the gSF approach for material interpolation, wherein Young's modulus of an element is interpolated as~\cite{sarkar2026generalized}:
\begin{equation} \label{Material_interpolation}
	E_e = E_\text{min}+\sum_{m=1}^{2^n} (\bar{\rho}_{me})^p(E_m-E_\text{min}).
\end{equation}
Here, $E_e$ is Young's modulus of element $e$, and $E_m$ is Young's modulus of material $m$ within that element. $E_\text{min}$ denotes a very small value of Young's modulus assigned to void material to avoid singularities in the stiffness matrix. $\bar{\rho}_{me}$ denotes the physical density of material $m$ within element $e$. $p$ is the penalization factor, set to 3.

The objective function minimized in this work is structural compliance with given volume constraints. Mathematically, the optimization problem is written as:
\begin{equation} \label{Optimization_eqn.}
    \begin{split}
        & \min:\quad f_0 = \mathbf{F}^\top \mathbf{U} \\
        &\text{subjected to:}\\
         &\bm{\lambda}:\,\,\mathbf{K} \mathbf{U} = \mathbf{F} \,\,
        \bm{\mu}:\,\,g_m=\frac{V_{m}^\mathrm{t}}{V_{m}^\mathrm{a}}-1 \le 0 \quad(V_{m}^\mathrm{t}=\sum_{e=1}^{N}v_{e}\bar{\rho}_{me}),\\
        &\quad\,\,\,\, -1 \leq \zeta_{1e},...,\zeta_{ne} \leq 1 \quad(e=1,\,2,...,\,N),
    \end{split}   
\end{equation}
where $f_0$ denotes the compliance of the structure, $v_e$ is the volume of element $e$, and $\bar{\rho}_{me}$ represents the physical density of material $m$ within element $e$. $V_{m}^\mathrm{a}$ denotes the total allowable volume for material $m$, while $V_{m}^\mathrm{t}$ represents the total volume of material $m$ in the current design domain. $\mathbf{F}$ and $\mathbf{U}$ denote the global force and displacement vector, respectively. $\mathbf{K}$ represents the global stiffness matrix. $g_m$ denotes the volume constraint for material $m$ $(m=1,\,2,...,\,M)$, and $M$ is the number of materials used for the optimization. $\zeta_{ne}$ is the $n$-th design variable of element $e$. The vectors $\bm{\lambda}$ and $\bm{\mu}$ denote the Lagrange multipliers corresponding to the respective constraints. A gradient-based optimization algorithm, the Method of Moving Asymptotes~\cite{svanberg1987method}, is employed to update the design variables.

\section{Results and Discussions}\label{sec4}
\begin{figure}[]
	\centering
	\begin{subfigure}{0.45\textwidth}
		\centering
		\includegraphics[width=0.8\textwidth]{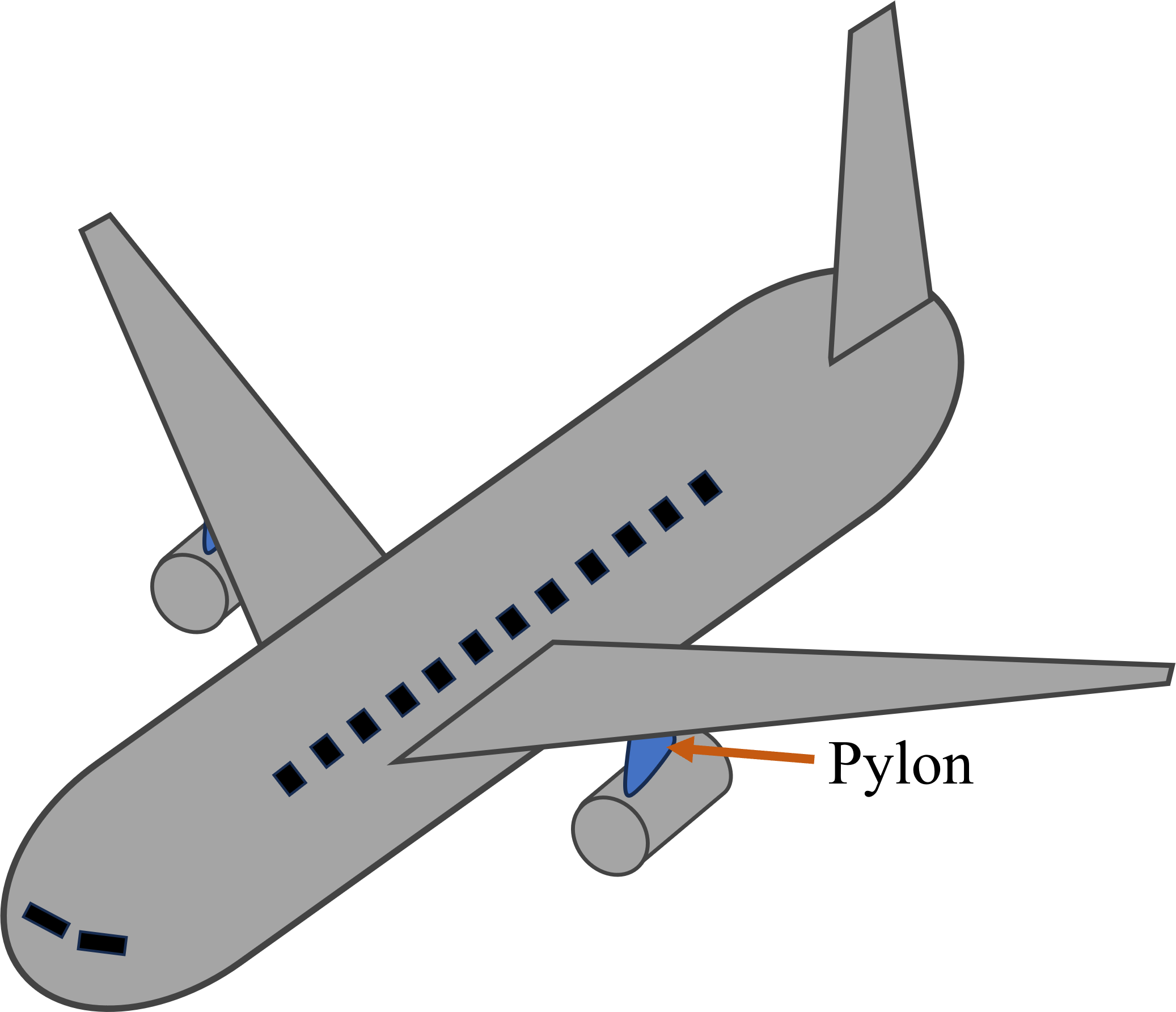}
		\caption{Schematic of an aircraft pylon}
		\label{Pylon_2D_schematic}
	\end{subfigure}
	\begin{subfigure}{0.46\textwidth}
		\centering
		\begin{tikzpicture}[scale=0.45]
			\makeatletter
			\pgfdeclarepatternformonly[\LineSpace,\LineThickness]{custom hatch}
			{\pgfqpoint{0pt}{0pt}}
			{\pgfqpoint{\LineSpace}{\LineSpace}}
			{\pgfqpoint{\LineSpace}{\LineSpace}}
			{
				\pgfsetlinewidth{\LineThickness}
				\pgfpathmoveto{\pgfqpoint{0pt}{0pt}}
				\pgfpathlineto{\pgfqpoint{\LineSpace}{\LineSpace}}
				\pgfusepath{stroke}
			}
			\makeatother
			
			\tikzset{
				LineSpace/.store in=\LineSpace,
				LineSpace=2pt,
				LineThickness/.store in=\LineThickness,
				LineThickness=0.4pt
			}
			
			\fill[pattern=custom hatch] (6.3,4.5) rectangle (9.0,5.0);
			\fill[gray!50] (0,0) rectangle (9,4.5);
			
			\draw[black, very thick] (0,0) -- (9,0);
			\draw[black, very thick] (9,0) -- (9,4.5);
			\draw[black, very thick] (9,4.5) -- (0,4.5);
			\draw[black, very thick] (0,4.5) -- (0,0);
			
			\dimline[
				color=blue,
				extension start length=2pt,
				extension end length=2pt,
				label style={fill=none, yshift=6pt}
			]{(0,6)}{(9,6)}{$L_x$};
			
			\dimline[
				color=blue,
				extension start length=2pt,
				extension end length=2pt,
				label style={fill=none, yshift=6pt}
			]{(6.3,5.2)}{(9,5.2)}{$0.3L_x$};
			
			\draw[red, very thick, -Stealth] (0,0) -- (0,-1.5);
			\draw[red, very thick, -Stealth] (1.5,0) -- (1.5,-1.5);
			\draw[red, very thick, -Stealth] (3.0,0) -- (3.0,-1.5);
			\draw[red, very thick, -Stealth] (4.5,0) -- (4.5,-1.5);
			\draw[red, very thick, -Stealth] (6.0,0) -- (6.0,-1.5);
			\draw[red, very thick, -Stealth] (7.5,0) -- (7.5,-1.5);
			\draw[red, very thick, -Stealth] (9,0) -- (9,-1.5);
			
			\dimline[
				color=blue,
				extension start length=2pt,
				extension end length=2pt,
				label style={fill=none, xshift=4pt, yshift=6pt}
			]{(9.4,4.5)}{(9.4,0)}{$L_y$};
			
			\node at (3.75,-1.0) {$F$};
			
		\end{tikzpicture}
		\caption{2D design domain of a pylon}
		\label{Pylon_2D}
	\end{subfigure}
	
	\caption{Schematic representations of an aircraft pylon and its corresponding 2D design domain.}
	\label{Pylon_2D_schematics}
	
\end{figure}
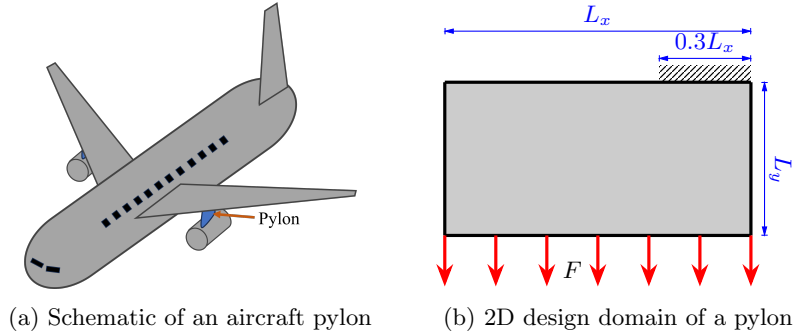

This section provides the optimized aircraft pylon with 14 different materials. A schematic diagram of a pylon is shown in Fig.~\ref{Pylon_2D_schematic}. Using the geometrical features of the pylon structure, the midplane is selected for optimization; its design domain, with the boundary and load, is shown in Fig.~\ref{Pylon_2D}. $L_x$ and $L_y$ represent dimensions along the $x$- and $y$-axes, respectively. A portion of the top edge, spanning 30\% from its right end, is held fixed, while a constant distributed force $F$ acts along the bottom edge. Herein, we consider only the force due to the fuselage's weight.

The optimized pylon, up to 14 materials, is presented. We use 2D shape functions (SFs) material element for $2$ and $3$ material TO, 3D SFs for $4{-}7$ material TO, and 4D SFs for $8{-}14$ material TO. The 2D design domain is discretized with $150$ FEs along the $x$-direction and $60$ FEs along the $y$-direction. $r_\mathrm{min}$ is set to 6, and $\gamma$ is taken as $0.5$. $\beta$ is initially set to $1$ and doubled after every $50$ iterations, up to a maximum of $64$, to accommodate crisp solutions. The value of force is taken as $F=1$. The maximum allowable volume fraction is set to $0.2$, $0.08$, and $0.04$ for two-dimensional, three-dimensional, and four-dimensional SFs, respectively. Material properties are normalized such that the Young's modulus (${E}_{\mathrm{norm}}$) of the stiffest candidate material is set to unity, with the moduli of all remaining materials scaled proportionally relative to this reference value. 

\begin{table}
\caption{Optimized results (I)}\label{3Dpylon}
\begin{subtable}[t]{1\textwidth}
\centering
\begin{tabular}{|C{1.0cm}|C{4.2cm}C{4.2cm}|C{2.0cm}|}
\hline
$M$ & \makecell{2D optimized design} & \makecell{3D optimized design} & \makecell{Output data} \\ \hline
$2$ &  \includegraphics[scale=0.35]{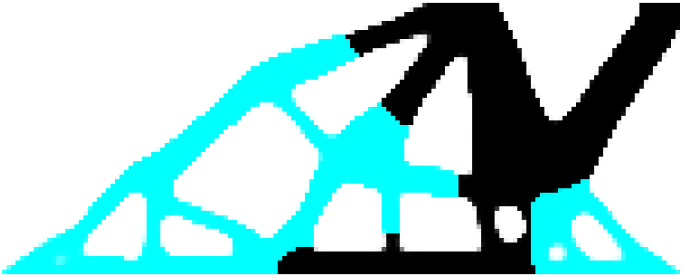} &  \vspace{0.1mm} \includegraphics[scale=0.35]{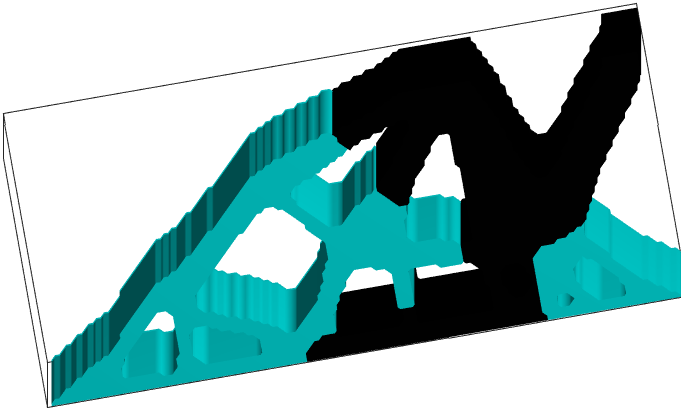} & \makecell{$f_0=0.5765$ \\ $M_{\mathrm{nd}}=1.70\%$} \\ \hline
$3$ &  \includegraphics[scale=0.35]{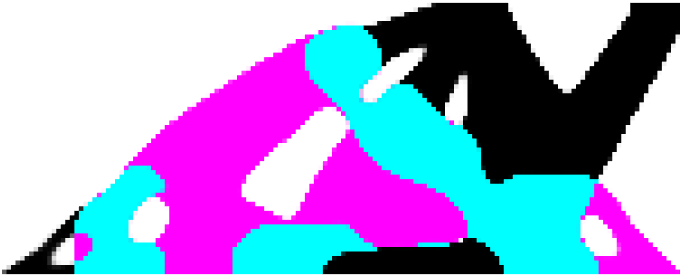} &  \vspace{0.1mm} \includegraphics[scale=0.35]{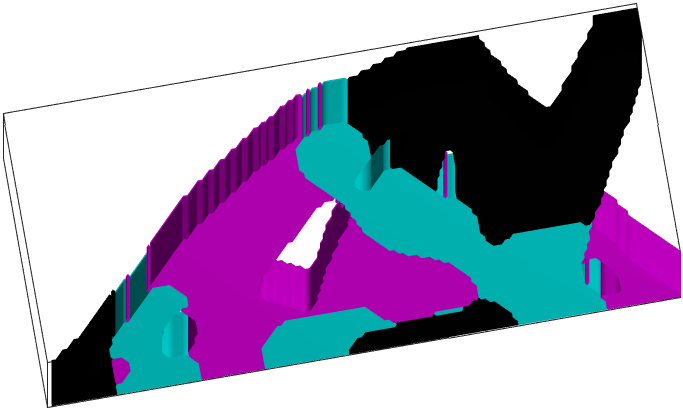} & \makecell{$f_0=0.5391$ \\ $M_{\mathrm{nd}}=0.85\%$} \\ \hline
$4$ &  \includegraphics[scale=0.35]{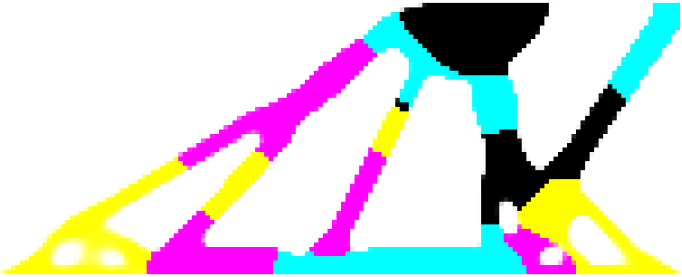} &  \vspace{0.1mm} \includegraphics[scale=0.35]{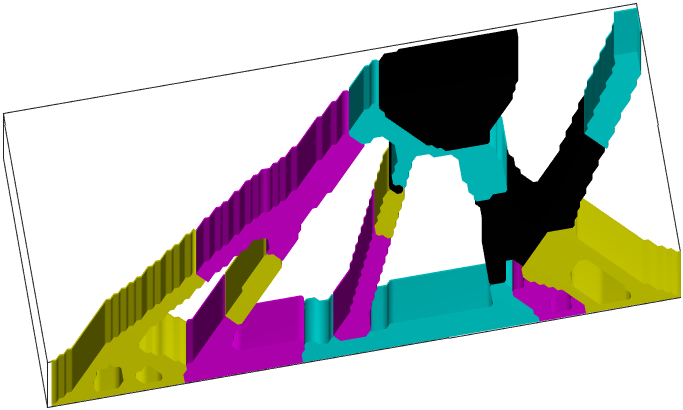} & \makecell{$f_0=0.1987$ \\ $M_{\mathrm{nd}}=2.84\%$} \\ \hline
$5$ &  \includegraphics[scale=0.35]{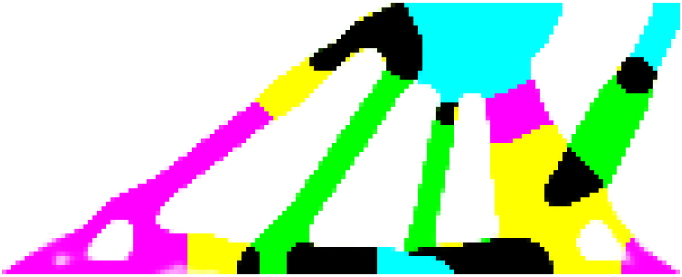} &  \vspace{0.1mm} \includegraphics[scale=0.35]{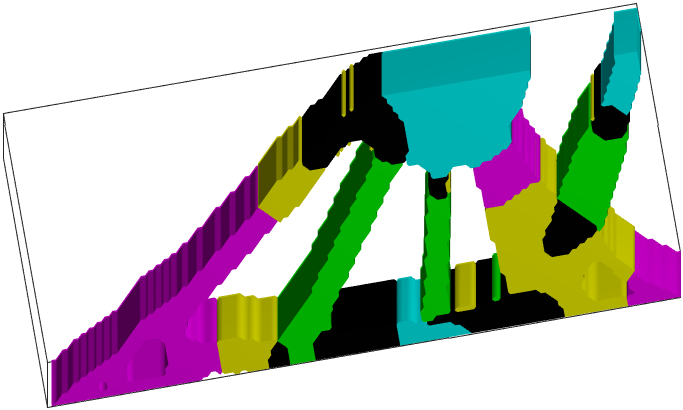} & \makecell{$f_0=0.1927$ \\ $M_{\mathrm{nd}}=1.76\%$} \\ \hline
$6$ &  \includegraphics[scale=0.35]{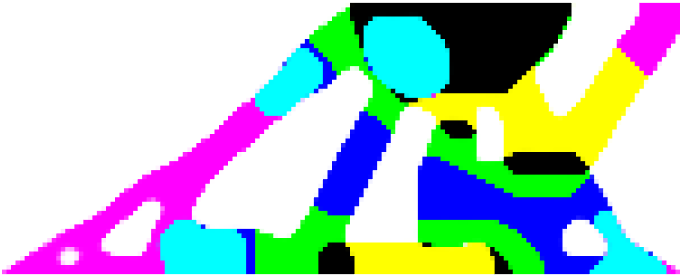} &  \vspace{0.1mm} \includegraphics[scale=0.35]{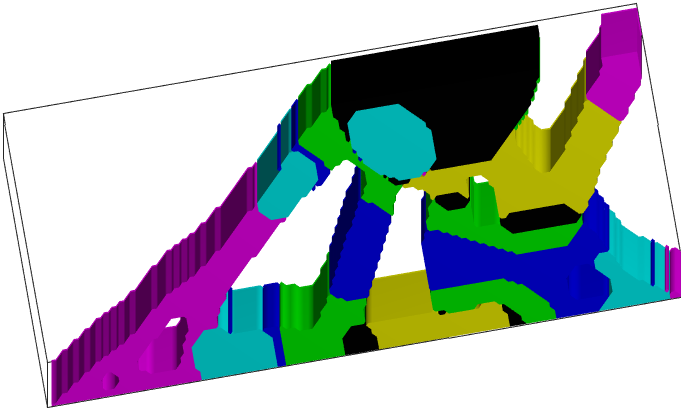} & \makecell{$f_0=0.1883$ \\ $M_{\mathrm{nd}}=1.04\%$} \\ \hline
$7$ &  \includegraphics[scale=0.35]{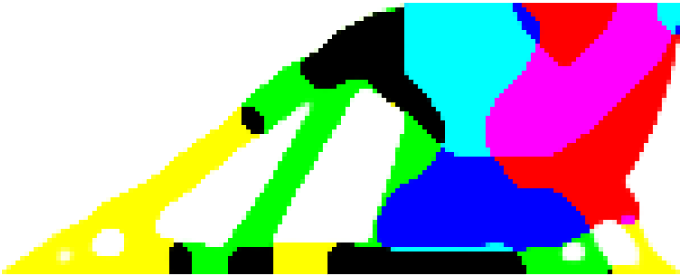} &  \vspace{0.1mm} \includegraphics[scale=0.35]{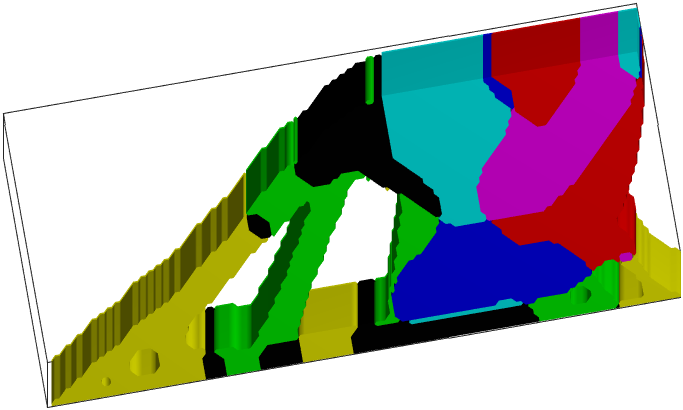} & \makecell{$f_0=0.1770$ \\ $M_{\mathrm{nd}}=1.21\%$} \\ \hline
\end{tabular}
\caption{Optimized designs with $2,\cdots,7$ materials}\label{2to7M}
\end{subtable}

\begin{subtable}[t]{0.46\textwidth}
    \centering
    \begin{tabular}{|c|c|c|c|}
        \hline
        \text{Colour} & $E_{\mathrm{norm}}$ &
        \text{Colour} & $E_{\mathrm{norm}}$ \\ \hline
        \cellcolor{white} & $10^{-9}$ &
        \cellcolor{magenta} & $1/3$ \\ \hline
        \cellcolor{cyan} & $2/3$ &
        \cellcolor{black} & $1$ \\ \hline
    \end{tabular}
    \subcaption{Colour schemes for $M=2,3$}
    \label{M23}
\end{subtable}
\hfill
\begin{subtable}[t]{0.46\textwidth}
    \centering
    \begin{tabular}{|c|c||c|c||c|c||c|c|}
        \hline
        \text{Colour} & $E_{\mathrm{norm}}$ &
        \text{Colour} & $E_{\mathrm{norm}}$ &
        \text{Colour} & $E_{\mathrm{norm}}$ &
        \text{Colour} & $E_{\mathrm{norm}}$ \\ \hline
        \cellcolor{white} & $10^{-9}$ &
        \cellcolor{red} & $1/7$ &
        \cellcolor{blue} & $2/7$ &
        \cellcolor{green} & $3/7$ \\ \hline
        \cellcolor{yellow} & $4/7$ &
        \cellcolor{magenta} & $5/7$ &
        \cellcolor{cyan} & $6/7$ &
        \cellcolor{black} & $1$ \\ \hline
    \end{tabular}
    \subcaption{Colour schemes for $M=4,5,6,7$}
    \label{M47}
\end{subtable}

\end{table}

\begin{table}
\caption{Optimized results (II)}\label{4Dpylon}
\begin{subtable}[t]{1\textwidth}
\centering
\begin{tabular}{|C{1.0cm}|C{4.2cm}C{4.2cm}|C{2.0cm}|}
\hline
$M$ & \makecell{2D optimized design} & \makecell{3D optimized design} & \makecell{Output data} \\ \hline
$8$ &  \includegraphics[scale=0.35]{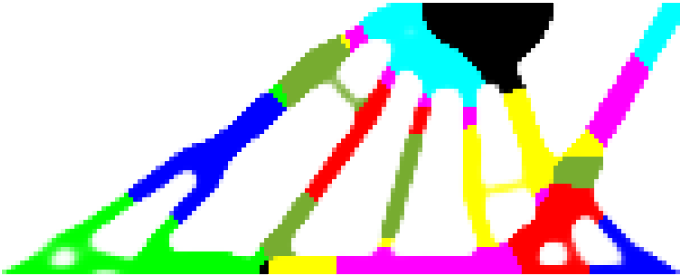} &  \vspace{0.1mm} \includegraphics[scale=0.35]{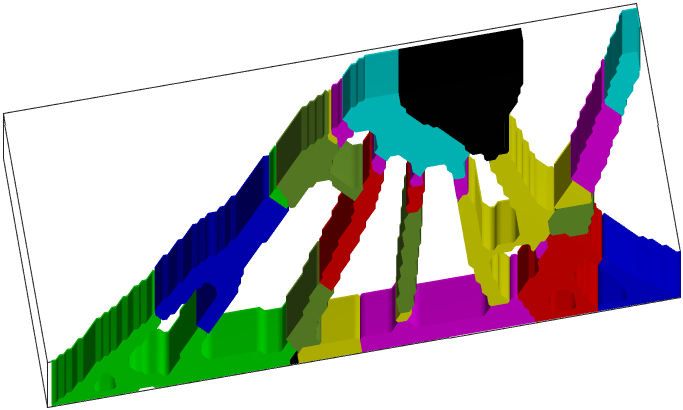} & \makecell{$f_0=0.0486$ \\ $M_{\mathrm{nd}}=3.78\%$} \\ \hline
$10$ &  \includegraphics[scale=0.35]{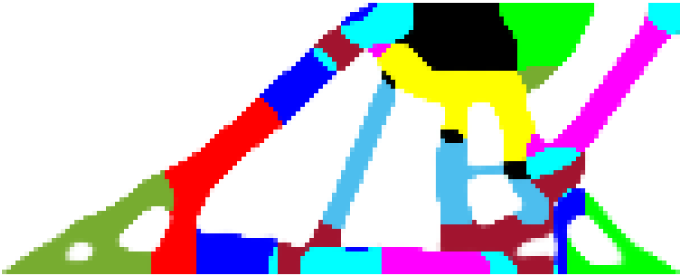} &  \vspace{0.1mm} \includegraphics[scale=0.35]{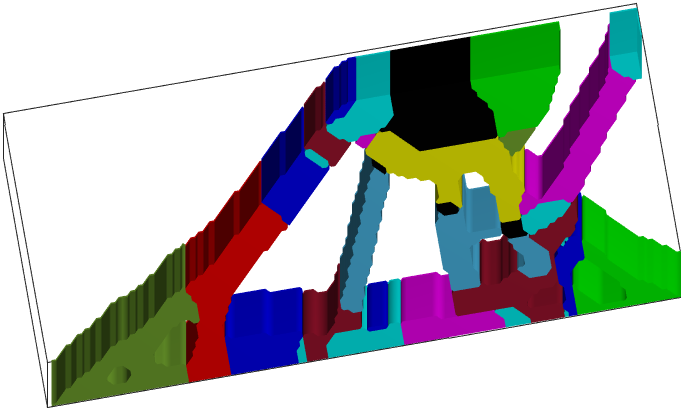} & \makecell{$f_0=0.0474$ \\ $M_{\mathrm{nd}}=1.90\%$} \\ \hline
$12$ &  \includegraphics[scale=0.35]{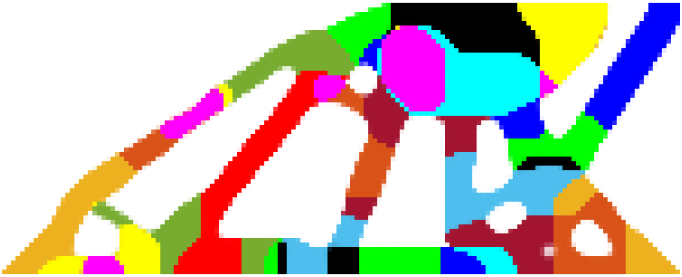} &  \vspace{0.1mm} \includegraphics[scale=0.35]{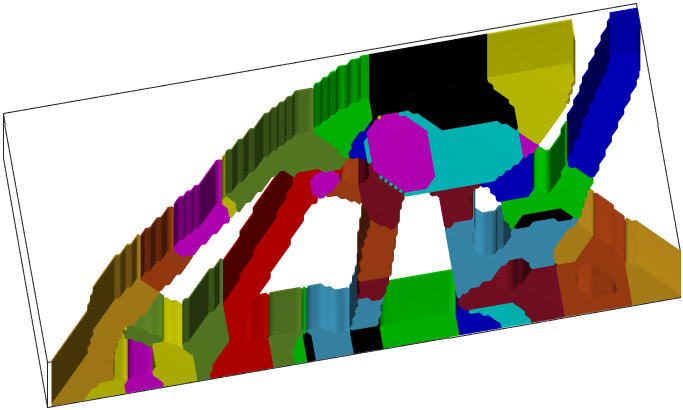} & \makecell{$f_0=0.0460$ \\ $M_{\mathrm{nd}}=2.04\%$} \\ \hline
$14$ &  \includegraphics[scale=0.35]{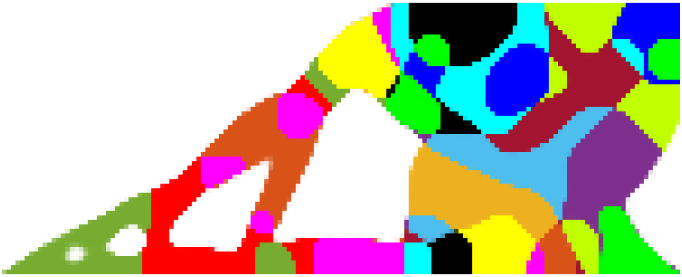} &  \vspace{0.1mm} \includegraphics[scale=0.35]{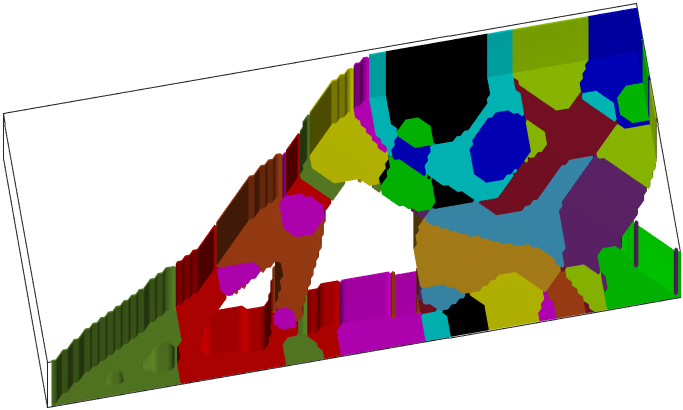} & \makecell{$f_0=0.0464$ \\ $M_{\mathrm{nd}}=1.11\%$} \\ \hline
\end{tabular}
\caption{Optimized designs with $8$, $10$, $12$, and $14$ materials}\label{8101214M}
\end{subtable}

\begin{subtable}[t]{\textwidth}
    \centering
    \begin{tabular}{|c|c||c|c||c|c||c|c||c|c||c|c|}
        \hline
        \text{Colour} & $E_{\mathrm{norm}}$ &
        \text{Colour} & $E_{\mathrm{norm}}$ &
        \text{Colour} & $E_{\mathrm{norm}}$ &
        \text{Colour} & $E_{\mathrm{norm}}$ &
        \text{Colour} & $E_{\mathrm{norm}}$ &
        \text{Colour} & $E_{\mathrm{norm}}$ \\ \hline

        \cellcolor{white}       & $10^{-9}$ &
        \cellcolor{olivegreen}  & $1/15$ &
        \cellcolor{greenyellow} & $2/15$ &
        \cellcolor{violet}      & $3/15$ &
        \cellcolor{yellowochre} & $4/15$ &
        \cellcolor{redorange}   & $5/15$ \\ \hline

        \cellcolor{brownred}    & $6/15$ &
        \cellcolor{skyblue}     & $7/15$ &
        \cellcolor{shafgreen}   & $8/15$ &
        \cellcolor{red}         & $9/15$ &
        \cellcolor{blue}        & $10/15$ &
        \cellcolor{green}       & $11/15$ \\ \hline

        \cellcolor{yellow}      & $12/15$ &
        \cellcolor{magenta}     & $13/15$ &
        \cellcolor{cyan}        & $14/15$ &
        \cellcolor{black}       & $1$ &
        & & & \\ \hline
    \end{tabular}
    \caption{Colour schemes for $M=8,\dots,15$}
    \label{M815}
\end{subtable}

 \end{table}

\noindent For instance, in the two-material problem, the stiffest material is assigned $E_{\mathrm{norm}}=1$, while the other material is assigned $E_{\mathrm{norm}}=2/3$. Similarly, in the three-material problem, the stiffest material retains $E_{\mathrm{norm}}=1$, the intermediate material is assigned $E_{\mathrm{norm}}=2/3$, and the softest material is assigned $E_{\mathrm{norm}}=1/3$. This proportional scaling of normalized Young's moduli, from the stiffest to the softest candidate material, is applied consistently across all material sets considered in this work. 
\begin{figure}
\centering
	\begin{tikzpicture}[scale=0.85][remember picture]
			\pgfplotsset{compat=1.9}
			\begin{axis}[
				width = 1\textwidth,
				xlabel=  Iteration number,
				ylabel=  Objective function $(f_0)$,
				xmin=0,xmax=400,
                scaled y ticks=false, 
                y tick label style={/pgf/number format/fixed, /pgf/number format/precision=3},
				grid=both,
				major grid style={line width=0.2pt, draw=gray!30},
                legend columns=2,
                legend style={
                    /tikz/column 2/.style={
                        column sep=1em
                    }
                },]	
                \pgfplotstableread{Pylon2D_SF2D2M_vf20x2E1by3_150x60full_400iter_beta50.txt}\mydata;
				\addplot[smooth,{cyan}, line width=1pt, mark = none]
                table[x index=1, y index=3] {\mydata};
				\addlegendentry{$M=2$}			
				\pgfplotstableread{Pylon2D_SF2D3M_vf20x3E1by3_150x60full_400iter_beta50.txt}\mydata;
				\addplot[smooth,{magenta}, line width=1pt, mark = none]
                table[x index=1, y index=3] {\mydata};
				\addlegendentry{$M=3$}	
				\pgfplotstableread{Pylon2D_SF3D4M_vf08x4E1by7_150x60full_400iter_beta50.txt}\mydata;
				\addplot[smooth,{yellow}, line width=1pt, mark = none]
                table[x index=1, y index=3] {\mydata};
				\addlegendentry{$M=4$}			
				\pgfplotstableread{Pylon2D_SF3D5M_vf08x5E1by7_150x60full_400iter_beta50.txt}\mydata;
				\addplot[smooth,{green}, line width=1pt, mark = none]
                table[x index=1, y index=3] {\mydata};
				\addlegendentry{$M=5$}			
				\pgfplotstableread{Pylon2D_SF3D6M_vf08x6E1by7_150x60full_400iter_beta50.txt}\mydata;
				\addplot[smooth,{blue}, line width=1pt, mark = none]
                table[x index=1, y index=3] {\mydata};
				\addlegendentry{$M=6$}			
				\pgfplotstableread{Pylon2D_SF3D7M_vf08x7E1by7_150x60full_400iter_beta50.txt}\mydata;
				\addplot[smooth,{red}, line width=1pt, mark = none]
                table[x index=1, y index=3] {\mydata};
				\addlegendentry{$M=7$}
				\pgfplotstableread{Pylon2D_SF4D8M_vf04x8E1by15_150x60full_400iter_beta50.txt}\mydata;
				\addplot[smooth,{shafgreen}, line width=1pt, mark = none]
                table[x index=1, y index=3] {\mydata};
				\addlegendentry{$M=8$}
				\pgfplotstableread{Pylon2D_SF4D10M_vf04x10E1by15_150x60full_400iter_beta50.txt}\mydata;
				\addplot[smooth,{brownred}, line width=1pt, mark = none]
                table[x index=1, y index=3] {\mydata};
				\addlegendentry{$M=10$}
				\pgfplotstableread{Pylon2D_SF4D12M_vf04x12E1by15_150x60full_400iter_beta50.txt}\mydata;
				\addplot[smooth,{yellowochre}, line width=1pt, mark = none]
                table[x index=1, y index=3] {\mydata};
				\addlegendentry{$M=12$}
				\pgfplotstableread{Pylon2D_SF4D14M_vf04x14E1by15_150x60full_400iter_beta50.txt}\mydata;
				\addplot[smooth,{yellowochre}, line width=1pt, mark = none]
                table[x index=1, y index=3] {\mydata};
				\addlegendentry{$M=14$}			
			\end{axis}
		\end{tikzpicture}
        \caption{Objective convergence plots}
        \label{convergence}
\end{figure}
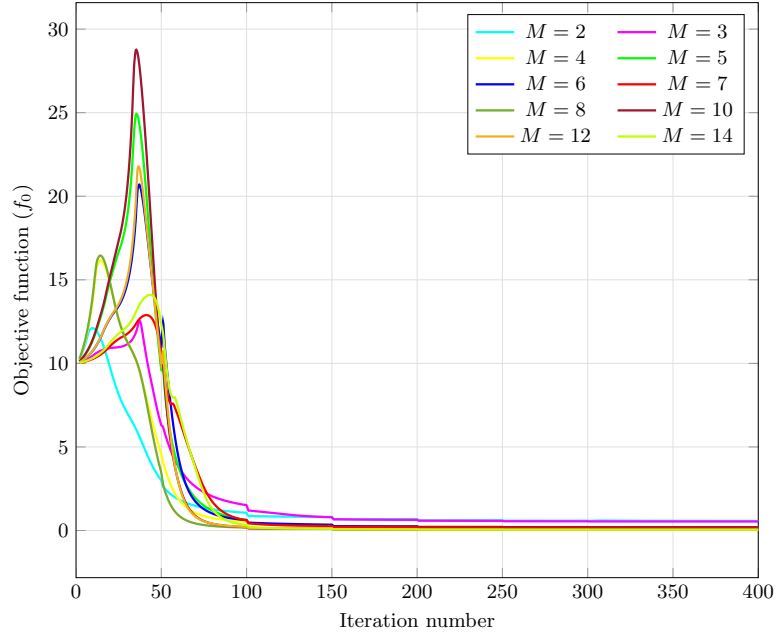
\noindent Void elements are assigned a Young's modulus of $10^{-9}$. The correspondence between color schemes and Young's moduli for each material is summarized in Tables~\ref{M23},~\ref{M47}, and~\ref{M815}. Poisson's ratio is set to $0.3$ uniformly across all materials.

The pylon design in this study exhibits geometric and loading symmetry along the $z$-direction, such that the resulting optimized topology is expected to remain uniform along this axis. Exploiting this symmetry, the three-dimensional result is instead obtained by extruding the converged two-dimensional optimized solution along the $z$-direction, to a length equal to the domain's extent in the $y$-direction. This approach yields a three-dimensional representation of the design while substantially reducing computational cost, since the optimization is performed on a two-dimensional domain. The resulting three-dimensional topologies are presented side by side with their corresponding two-dimensional counterparts in Tables~\ref {2to7M} and~\ref {8101214M}, allowing direct visual comparison between the two representations.

The dimensionality of the shape functions is determined by the number of desired materials in each TO problem. The two and three materials employ the 2D material element (2D SFs); four to seven materials employ the 3D material element (3D SFs); and eight to fourteen materials employ the 4D material element (4D SFs). At each iteration, the objective function and its sensitivities are normalized using a multiplication factor $f_{\mathrm{norm}} = \frac{10}{f_0}\Big|_{\mathrm{iter}=1}$, where `$\mathrm{iter}$' denotes the current MMA iteration number. All design variables are initialized to zero, corresponding to equal initial densities across all candidate materials.

The results are optimized with many materials under the volume constraints. The $M_{\mathrm{nd}}$ values indicate the degree of discreteness of the optimized designs. The values are very close to zero, indicating a near-binary solution that achieves the pure-material states of each contributing material. The convergence histories of $f_0$, for all optimization cases are presented together in Fig.~\ref{convergence}. In all cases, $f_0$ reaches a nearly converged state well before the prescribed maximum of 400 iterations. Nevertheless, the optimization is continued until 400 iterations to allow the volume constraints to become fully active and to ensure consistent convergence of the material distributions.

\section{Conclusions}\label{sec5}
This paper presents optimized multimaterial pylons obtained via topology optimization using the generalized shape function (gSF) approach. The gSF method facilitates up to $2^n{-}1$ materials with just $n$ design variables per element, keeping the optimization tractable as material complexity grows. We use two-dimensional, three-dimensional, and four-dimensional material elements to optimize the pylon design, yielding structures that incorporate up to 14 distinct materials and demonstrate a multi-material pylon structure. Exploiting the symmetric nature of the pylon structure's geometry and static loading, the mid-plane is optimized. The optimized results are further extruded into their corresponding 3D optimized designs.

The presented numerical results are crisp and reach near $0{–}1$ material states, so that each candidate material in the final design approaches a pure physical phase rather than an intermediate mixture. This near-elimination of material blurring is incorporated into the design problem, resulting in smooth convergence of the objective values. At the final stage of optimization, the volume constraints are still active. Together, these outcomes demonstrate that the proposed formulation not only satisfies the prescribed volume constraints but also yields sharply defined, optimized design topologies.

We minimize the pylon's compliance under the prescribed loading conditions. The formulated algorithm, with the optimizer, strategically places each material where it contributes most to stiffness, converging toward a minimized compliance value while simultaneously keeping the volume fraction of each material at its maximum prescribed limit.


\section*{Acknowledgment}
PK thanks the Anusandhan National Research Foundation, India, for support under the MATRICS project, file number MTR/2023/000524. 


\end{document}